*Perspective*

# Renewable Energies in the Agricultural Sector: A Perspective Analysis of the Last Three Years

**Quetzalcoatl Hernandez-Escobedo [1,*], David Muñoz-Rodríguez [2], Alejandro Vargas-Casillas [3], José Manuel Juárez Lopez [1], Pilar Aparicio-Martínez [4], María Pilar Martínez-Jiménez [2] and Alberto-Jesus Perea-Moreno [2,*]**

1 Escuela Nacional de Estudios Superiores Unidad Juriquilla, Universidad Nacional Autonoma de Mexico, Queretaro 76210, Mexico; josemanueljuarezlopez@comunidad.unam.mx
2 Departamento de Física Aplicada, Radiología y Medicina Física, Campus Universitario de Rabanales, Universidad de Córdoba, 14071 Córdoba, Spain; qe2murod@uco.es (D.M.-R.); fa1majip@uco.es (M.P.M.-J.)
3 Engineering Institute Campus Juriquilla, Universidad Nacional Autónoma de Mexico, Queretaro 76210, Mexico; avargasc@iingen.unam.mx
4 Departamento de Enfermería, Farmacología y Fisioterapia, Campus de Menéndez Pidal, Universidad de Córdoba, 14071 Córdoba, Spain; n32apmap@uco.es
* Correspondence: g12pemoa@uco.es (A.-J.P.-M.); qhernandez@unam.mx (Q.H.-E.)

**Abstract:** Renewable energy arises as a tool for the supply of energy to the agriculture sector. Currently, there is a growing concern for the environment. This circumstance has led to technological progress in energy use in relation to natural resources and their availability for all productive sectors, including agriculture. The main objective of this work is to perform analysis from a bibliometric point of view and to analyze scientific advances in renewable energy and agriculture worldwide that have occurred in the last three years (2019–2021). The purpose of this study is to provide an overview of the last three years on the topic in order to contribute to the international scientific community, specifically towards collaboration between authors, institutions, and countries. A keyword analysis using community detection was applied to detect the five main clusters of this research and was largely dedicated to the following topics: renewable energy technologies in agriculture, bioenergy, sustainable agriculture, biomass energy, and the environmental impact of agriculture. The main countries found to be conducting research on renewable energy and agriculture include India, China, the United States, Italy, the United Kingdom, Poland, Indonesia, Germany, the Russian Federation, and Spain; the most important institutions conducting research in this area include the Ministry of Agriculture of the People's Republic of China, the Tashkent Institute of Irrigation and Agricultural Mechanization Engineers at the National Research University in Uzbekistan, the Chinese Academy of Agricultural Sciences, and the Grupo de Investigação em Engenharia e Computação Inteligente para a Inovação e o Desenvolvimento in Portugal. These results may contribute to the identification of new research needs and therefore to the development of future directions of research on renewable energies in the agricultural sector.

**Keywords:** renewable energy; agriculture; Scopus; bibliometric; sustainable development

## 1. Introduction

In 2015, the United Nations (UN) approved the 2030 Agenda for Sustainable Development. To achieve an environmental paradigm shift, the 2030 Agenda identified energy sustainability as a key element for ensuring the viability of the global economic system. The UN's aims are conveyed through 17 Sustainable Development Goals [1]. Goal seven refers to energy sustainability, which aims to "ensure access to affordable, secure, sustainable and modern energy".

For its part, the European Union, in June 2021, passed the European Climate Law. This law aims to encourage members of the European Union to achieve climate neutrality (negative emissions) by 2050 and sets a binding target for reduced net greenhouse gas emissions, with the target being an at least 55% reduction by 2030 compared to 1990 levels [2].

The relationship between renewable energy and agriculture will be one of the most important objectives of nations, and this relationship can currently be observed in smart and solar greenhouse covers [3]. The environmental impacts of human activities on rivers and agricultural crops (among others) will have to be addressed by governments, society, and universities. Energy has become the currency of the economy of nature [4]. These activities represented 75% of land use modification worldwide [5]. Crops have different resource requirements, are associated with disparate agricultural practices, and are cultivated at different times throughout the year [6], and according to FAOSTAT [7] there are 13,796,719,205 hectares of area harvested worldwide. This development has been one of the most aggressive activities driving deforestation, thus impacting biodiversity conservation [8]. A study conducted by Kissinger et al. [9] established that about three quarters of the world's forests have disappeared due to agricultural expansion practices, and the WWF [10] calculates an annual forest loss of 18.2 million acres. This relationship between agriculture and renewable energy can also be observed in Australia, where a study was conducted between sugar cane and energy demand; it was found that solar photovoltaic energy could help mitigate the consumption of electricity from fossil energy sources [11].

The use of energy in agricultural activity, especially electricity, is essential for its proper development. In fact, it is used in many different ways: in draining and pumping water, to operate the necessary machinery, in heating greenhouses, in the transport of products both within the farms themselves and to the centers where they are marketed, and in the transformation and conservation of foodstuffs [12–15]. Khan et al. [16], in order to promote the correct use of resources, evaluated the water needs of cotton and sorghum wheat crops grown in Sudan. They achieved both the correct planning and use of water and the correct dimensioning of photovoltaic panels that supplied energy to a pumping system that allowed water to be extracted from aquifers. Riahi et al. [17], aware of the climatic problems that affect agriculture, modeled and implemented a greenhouse that works with renewable energy sources. The system used photovoltaic and wind energy, which was supported by a battery, to power a ventilation and air conditioning system that managed the temperature inside the greenhouse at different times of the year. Nordberg et al. [18] highlighted the problem of different land uses and their competition. The authors, looking for a higher yield of resources, proposed shared use of the land for the installation of solar farms that produce renewable energy and the traditional uses of pasture and cultivation. This plan favors the conservation and biodiversity of the area while at the same time boosting commercial activity. The above study is an example of a positive relationship, but Terneus and Viteri [19] showed a different situation by analyzing the production of biofuel and ethanol and their influence on agriculture, water consumption, pollution, and the importing of petroleum products in Ecuador. The authors argue that the production of biofuel reduces the importing of petroleum products by 2%. In contrast, biofuel negatively influences both food production and water pollution. The authors conclude that the problem of malnutrition exists in regions of Ecuador where an increased percentage of land is used for biofuel production.

The coexistence between agriculture and renewable installations is becoming increasingly widespread and is moving towards a sustainable relationship with mutual benefits for both sectors. Scientific advances have made it possible to develop agrovoltaic projects, which generate spaces of coexistence between energy production, the promotion of the local economy, and the protection of biodiversity [14–16]. Weselek et al. [20] studied agrovoltaic systems and combined the production of renewable energy and food for the particular case of celeriac (Apium graveolens L. var. rapaceum). They highlighted that the percentage of radiation measured in the area destined for the agrovoltaic system decreased compared to the crop in full sun. They concluded that celeriac is a suitable crop for production in agrovoltaic systems as the harvestable yields were not significantly reduced and aerial biomass increased, which is positive as it is indicative of good root biomass. Weselek et al. [21] conducted another experiment and increased the number of crops to celeriac, winter wheat, potato, and clover–grass, measuring bioclimatic parameters such as soil temperature, photosynthetic active radiation, and the yields of each crop. Although they concluded that lower yields for the crops studied were likely under photovoltaic panels, they highlighted that under certain conditions, such as hot and dry climatic conditions, agrovoltaic systems could favor them. Sojun and Sumin [22] used an agrovoltaic system in the Republic of Korea and modeled the performance of the agro-photovoltaic system using two machine learning techniques: polynomial regression and deep learning. The authors also conducted a cost-benefit analysis concerning energy production and crop production that would allow stakeholders to calculate the return on investment.

Another important aspect linking renewable energies and agriculture is biomass production for energy purposes. According to data provided by the European Environment Agency (EEA), the first 6800 hectares of energy crops were planted in 2004, and the European Union is expected to produce 147 million ktoe by 2030 [23]. Ocak and Acar [24] studied the use of bio-waste of agricultural and animal origin in the production of electrical energy due to the importance of energy imports in Turkey, which mainly involve fossil fuels, thus raising concerns about the security of supply. The authors estimate that it is economically profitable to use waste, with some environmental problems also minimized compared to the use of non-renewable energy sources. Moustakas et al. [25] evaluated the potential of agricultural biomass in the western region of Greece as an energy source using the anaerobic digestion of plants. The authors estimated that 715,080 tonnes of biomass from sources such as herbaceous crops and tree pruning could be combined with the currently installed industrial capacity for anaerobic digestion to produce an estimated 775 GWh via electrical energy and 1119 GWh through thermal energy. Alanhoud et al. [26] considered livestock and poultry manure as a present and future problem in northern Jordan. Therefore, considering the possibility of its use as a biofuel source, they chemically and thermally characterized cow, sheep/goat, and chicken manure, chicken feathers, and poultry litter. The samples were subjected to pyrolysis processes, and they estimated that the biogas production capacity of the studied waste was 48.7 million $m^3$, which could be transformed into $10.1 \times 10^6$ GJ of energy and could therefore satisfy the energy needs of 5% of the population of the area from which the samples were obtained.

The main objective of this work is to analyze the scientific advances made in research on renewable energy and agriculture worldwide that have occurred in the last three years, thus contributing to the international scientific community by helping to determine both future trends and actual status and also aiding collaboration between authors, institutions, and countries.

## 2. Materials and Methods

The imprint of new information and communication technologies has marked the constant growth of information and knowledge. Thus, scientific and technological research and the dissemination of knowledge are essential activities for the satisfaction of growing social needs. However, in addition to producing and transferring knowledge, there is a need to evaluate this research process.

Peer review criteria have always been used to evaluate scientific production, but their limitations led to the development of metric studies of information that make it possible to explore, detect, and display relevant and significant information in large volumes of documents. This led to the emergence of a new discipline with a quantitative and objective approach, whose results constitute a useful source of information for evaluating scientific activity: bibliometrics.

Perea-Moreno et al. [27] established that the analysis of scientific publications is fundamental in the research process.

### *2.1. Data Processing*

Scopus is Elsevier's abstract and citation database, which was launched in 2004. Scopus covers nearly 36,377 titles (22,794 active titles and 13,583 inactive titles) from approximately 11,678 publishers, of which 34,346 are peer-reviewed journals in top-level subject fields such as the life sciences, social sciences, physical sciences, and health sciences. It covers three sources: book series, journals, and trade journals [28].

Browns et al. [29] indicate that a range of available bibliometric measures should be used collectively to yield a more comprehensive assessment of journal and article rankings rather than the use of a singular measure; however, Scopus is the most accurate scientific database.

Bibliometric analysis can be defined as the use of statistical and mathematical methods to map the bibliographic data available in different scientific publication databases so that the evolution of research trends and the contributions in a thematic area can be analyzed and evaluated [30,31].

In this study, the Scopus database was used to analyze worldwide advances in research on renewable energies in the agriculture sector over the last three years through bibliometric techniques. For this propose, a research equation was used in the Scopus database to obtain documents about renewable energy being applied in agriculture: (TITLE-ABS-KEY (renewable AND energ*) AND TITLE-ABS-KEY (agriculture)) AND PUBYEAR > 2018 AND PUBYEAR < 2022. In total, 1378 documents were obtained.

In Figure 1, the methodology used to develop the research is shown.

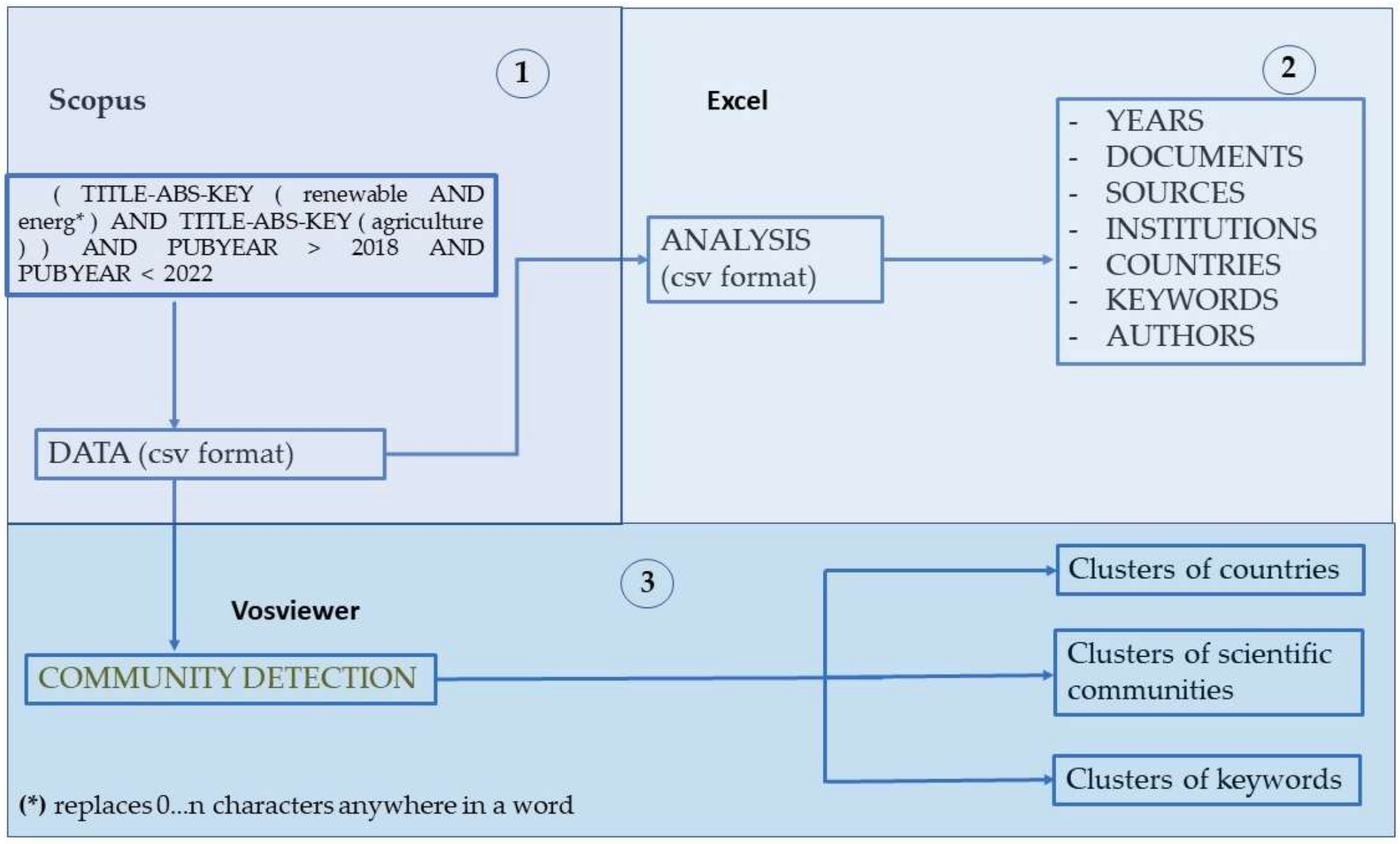

**Figure 1.** Methodology used.

Figure 1 presents the steps that were followed:

1. Worldwide search: Three fields were used in the Scopus search, i.e., (TITLE-ABS-KEY (renewable AND energ*) AND TITLE-ABS-KEY (agriculture)) AND PUBYEAR > 2018 AND PUBYEAR < 2022. The information was recorded as an excel file (.csv) and included information categories such as Years, Author, Document type, Cited by, Source, and Affiliation.
2. Analysis was performed separately, with each data point receiving its own analysis using excel software version 2211.

3. Clustering: information about authors, countries, and keywords was clustered using VOSviewer® software version 1.6.18, and the results provide information regarding scientific community/country collaborations and research trends through the use of keywords.

## 3. Results and Discussion

The number of documents by year, author, institution, subject, and country indicates the quantity of scientific activity [27]. If this information is compared to that of other institutions or countries, it can be useful and provide information about the scientific trends in the subject analyzed.

### *3.1. The Status of Research on Renewable Energy in the Agriculture Sector*

Energy use in agriculture is increasing, and with it the need to use green and sustainable energy. Scientific production regarding renewable energy and agriculture for the period 2019–2021 is shown in Figure 2.

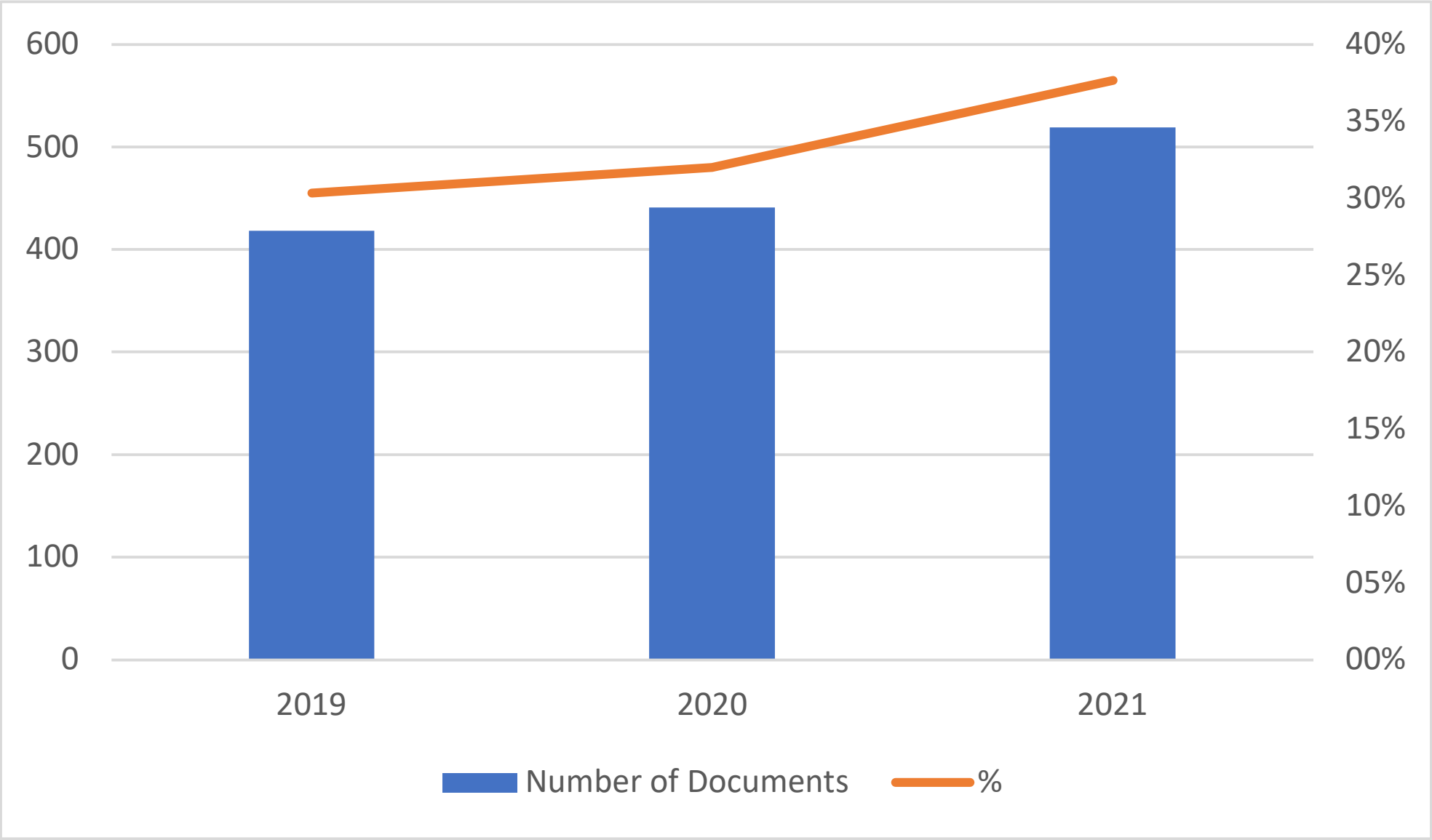

**Figure 2.** Renewable energy and agriculture publications for 2019–2021.

As can be seen, there is a clear increasing trend for research on the application of renewable energies in the agricultural sector, reaching maximum production (519 documents) and greatest increase (37.7%) in the year 2021.

There were 1378 documents from different types of publications, including articles (54.50%), conference papers (24.46%), reviews (11.54%), book chapters (6.31%), and books (0.87%), as well as other types of publication (2.32%) (Figure 3).

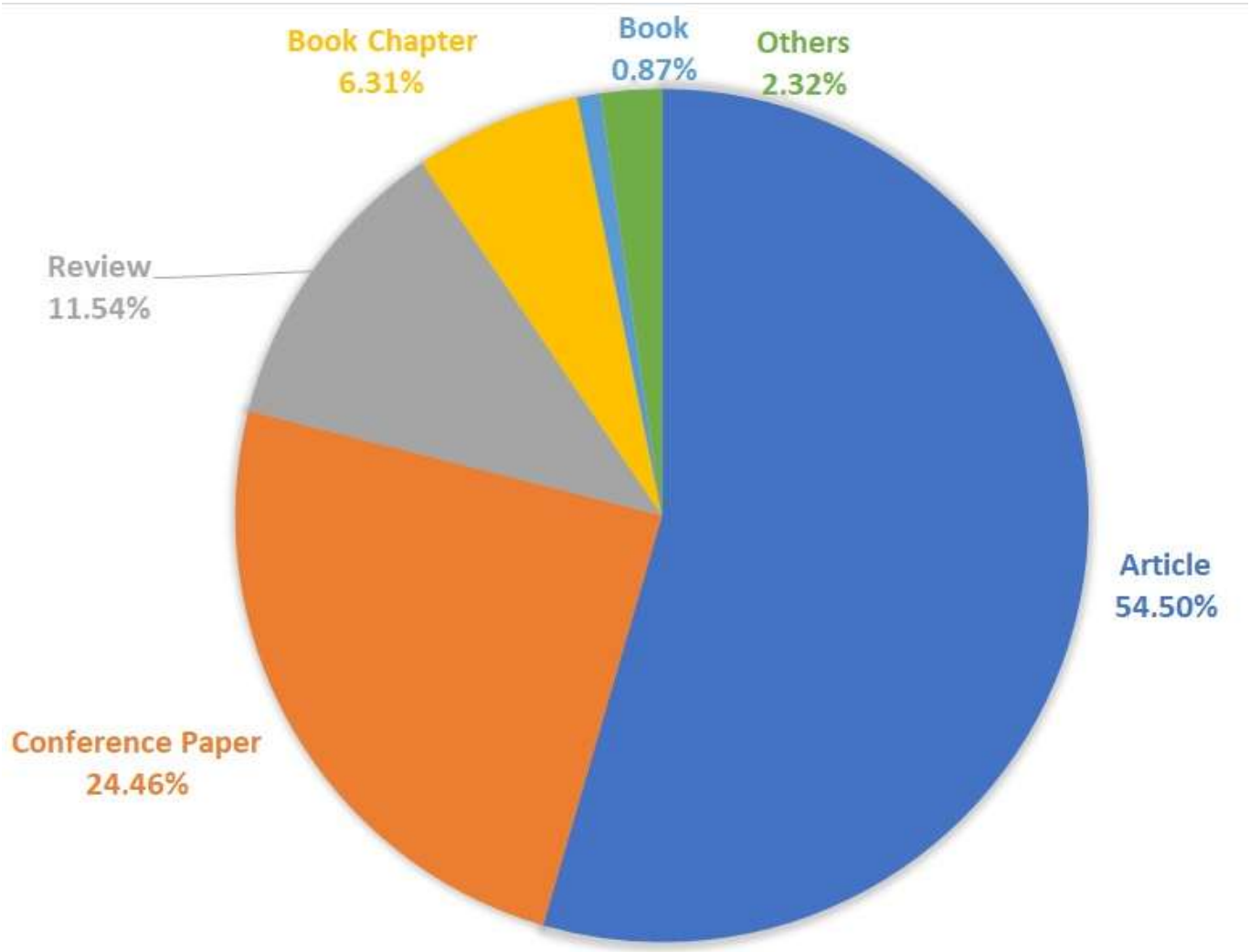

**Figure 3.** Renewable energy and agriculture document types for 2019–2021.

### *3.2. Worldwide Publications on Renewable Energy in the Agriculture Sector*

In Figure 4, global scientific production over the last three years is presented. The main countries with the most articles published were India (174 documents), China, the United States, Italy, the United Kingdom, Poland, Indonesia, Germany, the Russian Federation, and Spain with 12.63%, 11.97%, 11.32%, 6.68%, 5.37%, 5.15%, 4.86%, 4.50%, 4.14%, and 3.99% of publications, respectively.

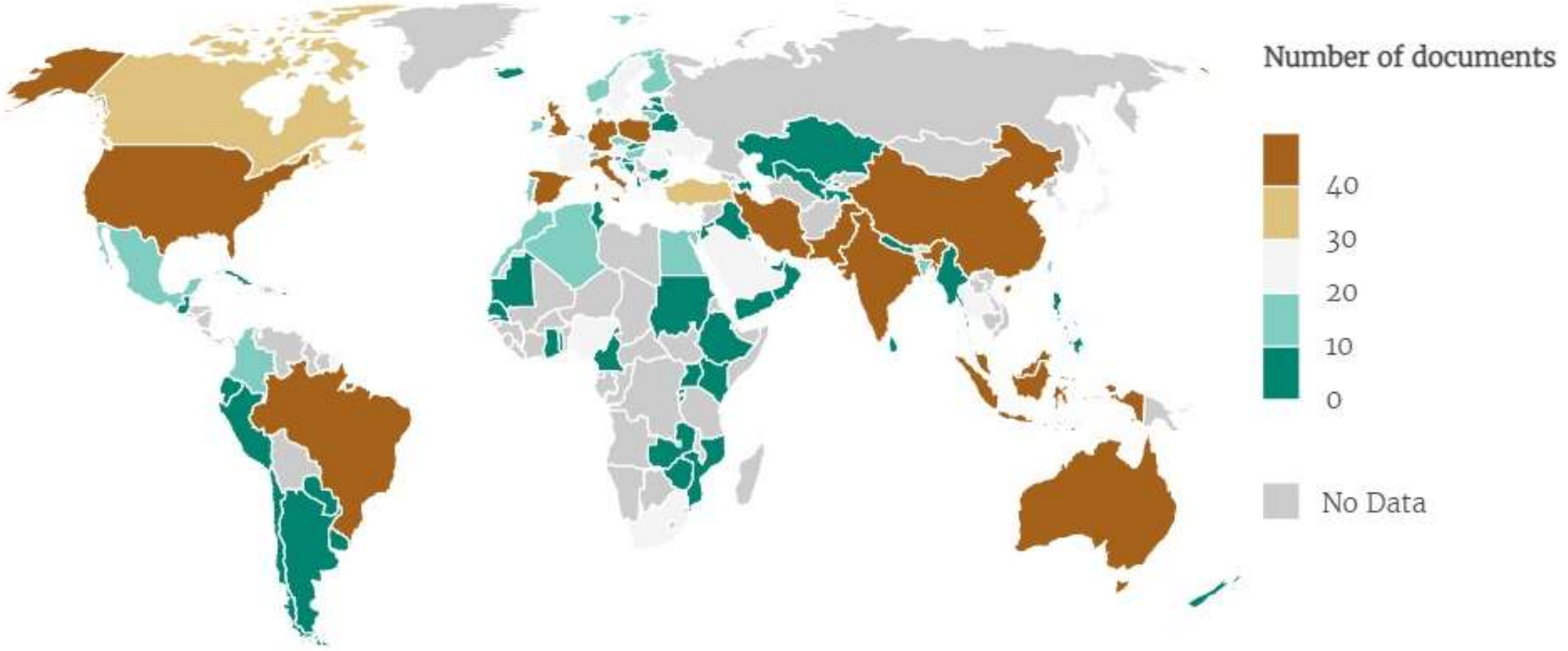

**Figure 4.** Worldwide scientific production related to renewable energy and agriculture (2019–2021).

The number of publications related to renewable energy and agriculture for each of the 10 main countries over the last three year is presented in Figure 5. It is worth noting the positive trend in all countries regarding research related to renewable energy and agriculture, though a decrease is observed in some countries in 2020 due to the COVID-19 pandemic. Some countries, such as the United States, Italy, the United Kingdom, Poland, Indonesia, and the Russian Federation, had a slower positive trend regarding the publication of studies in 2020 due to difficulties in carrying out research during lockdown periods.

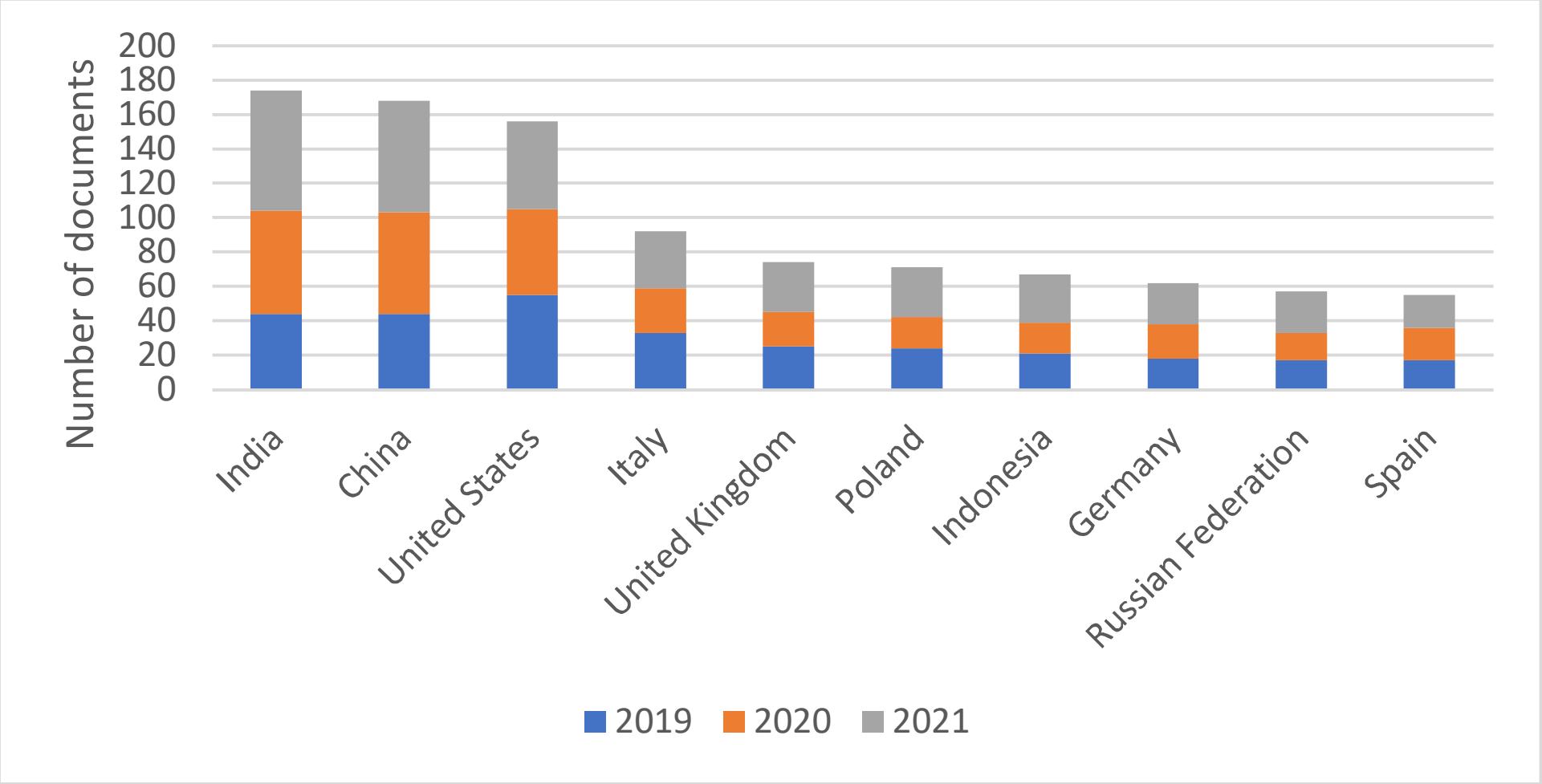

**Figure 5.** Evolutionary trends in renewable energy and agriculture publications from 2019 to 2021 for the top ten countries.

To determine the degree of collaboration between authors from different countries researching the same subject, VOSviewer® software version 1.6.18 was used. This application recorded the downloaded data from the Scopus database and the delivered *.csv file. A visualization of collaborative relationships can be seen in Figure 6. The color assigned to each country determines which cluster it belongs to. The size of the circle shows the importance of the country based on the number of publications on renewable energy and agriculture. The lines represent collaborations between the countries that form each cluster.

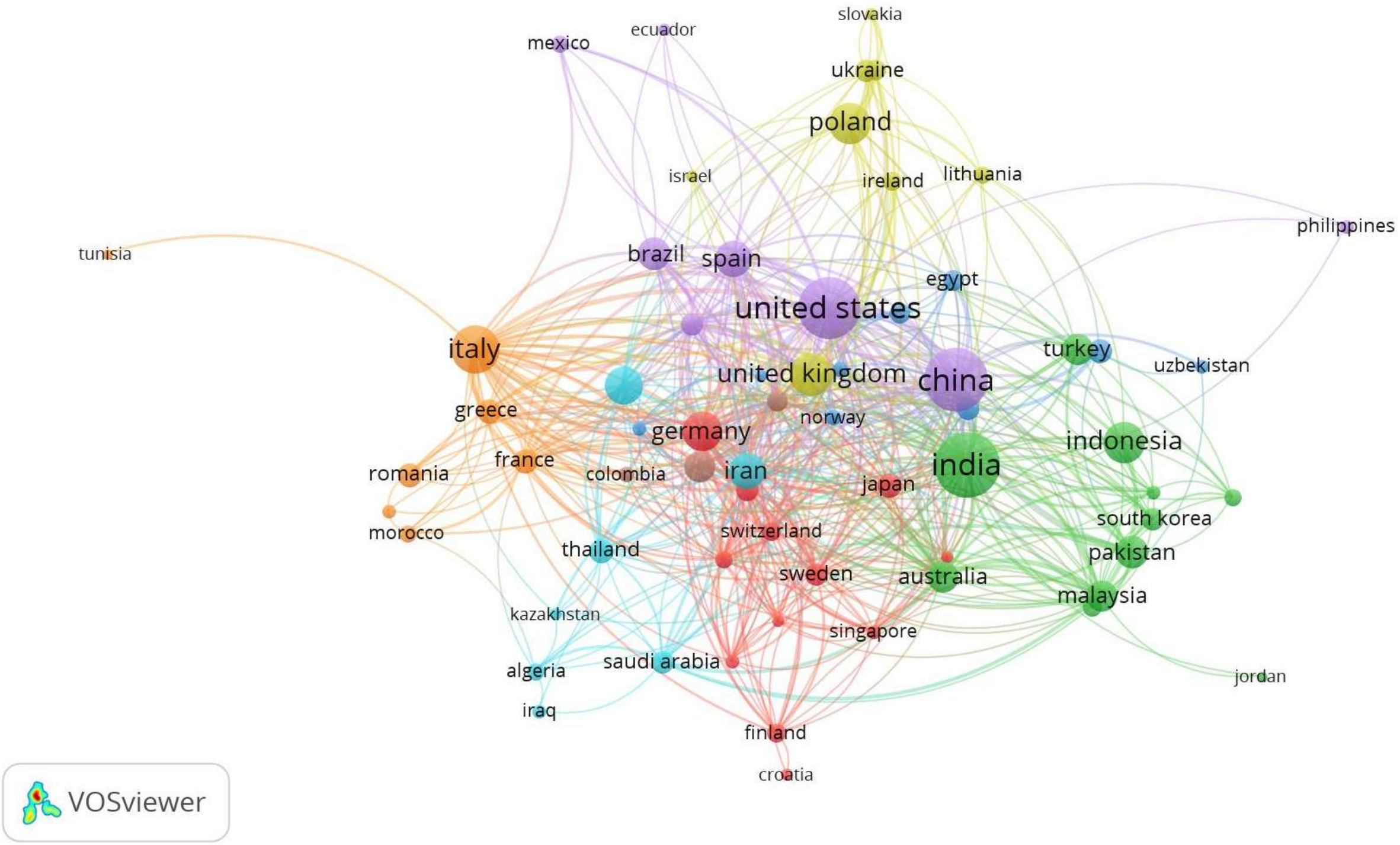

**Figure 6.** International collaboration clusters (2019–2021).

Eight collaboration clusters were calculated between countries regarding research on renewable energy and agriculture (Figure 6 and Table 1); however, India, the United States (US), the United Kingdom (UK), and Italy were more important countries based on their levels of collaboration. The red cluster has most collaborations, and these collaborations mainly occur between countries from Europe. The green cluster mainly consists of collaborations between Asian and Oceanian countries.

**Table 1.** Main clusters obtained in international collaboration analysis for the topic of renewable energies in agriculture.

| Cluster | Color | Countries |
|---|---|---|
| 1 | Red | Austria, Croatia, Denmark, Finland, Germany, Japan, Kenya, New Zealand, Singapore, Sweden, Switzerland, United Arab Emirates. |
| 2 | Green | Australia, Bangladesh, India, Indonesia, Jordan, Malaysia, Pakistan, South Korea, Taiwan, Turkey, Vietnam. |
| 3 | Blue | Chile, Egypt, Hungary, Nigeria, Norway, Portugal, Serbia, South Africa, Uzbekistan. |
| 4 | Yellow | Czech Republic, Ireland, Israel, Lithuania, Poland, Slovakia, Ukraine, United Kingdom. |
| 5 | Purple | Brazil, China, Ecuador, Mexico, Netherlands, Philippines, Spain, United States. |
| 6 | Turquoise | Algeria, Iran, Iraq, Kazakhstan, Russian Federation, Saudi Arabia, Thailand. |
| 7 | Orange | Bulgaria, France, Greece, Italy, Morocco, Romania, Tunisia. |
| 8 | Brown | Belgium, Canada, Colombia. |

*3.3. Worldwide Institution Distribution*

Knowing which institutions are leading research into renewable energy and agriculture will provide information about current trends and the most used keywords.

Table 2 presents the 10 most productive institutions in terms of research into renewable energy and agriculture, including the most used keywords.

**Table 2.** Ten most productive institutions in terms of research into renewable energy and agriculture (2019–2021).

| Institution | Documents | Country | Keywords Most Used | |
|---|---|---|---|---|
| | | | 1 | 2 |
| Chinese Ministry of Education | 18 | China | Agricultural Wastes | Biomass |
| University of Tehran | 13 | Iran | Agriculture | Agricultural Robots |
| China Agricultural University | 12 | China | Agricultural Robots | Sustainability |
| China Academy of Sciences | 12 | China | Alternative Energy | Agricultural Robots |
| Wageningen University and Research | 11 | Netherlands | Renewable Energy | Climate Change |
| Imperial College London | 11 | United Kingdom | Fossil Fuel | Agricultural Robots |
| Norwest A&F University | 10 | China | Agriculture Robots | Alternative Energy |
| Universidade de Sao Paulo | 9 | Brazil | Agriculture Robots | Agriculture |
| Universitat Autònoma de Barcelona | 9 | Spain | Agriculture | Renewable Energies |
| Uniwerytet Warmisko-Mazurski w Olsztynie | 9 | Poland | Agriculture Robots | Biomass |

In conclusion, it is worth highlighting the importance agriculture robots have acquired in recent years. There are a number of robots worldwide that are being tested in agricultural tasks, which is made possible by the use of GPS, the development of sensors and hardware, and the integration of renewable energies, all of which are used by the robots to position themselves in the field to perform their tasks [32–34]. Also noteworthy is the importance of the agricultural sector in the production of bioenergy from energy crops or waste from the agri-food industry. There are many studies that demonstrate the importance of this source of energy [35–38].

### 3.4. Distribution of Publications

The distribution of publications by thematic areas is presented in Figure 7. The illustration shows that the area with the highest number of documents is environmental science with 598, which is equivalent to 19.8% of documents published on renewable energy in this area; the next areas with the most documents are energy with 19.1%, engineering with 14.5%, agricultural and biological sciences with 6.7%, and earth and planetary sciences with 5.0%. The other areas, such as social sciences, computer science, mathematics, etc., represent 34.9% of the subject areas were publications were found.

An interesting methodology is that used by Perea-Moreno et al. [27], where they establish that assessment of the impact of a work according to its citations is not an immediate measure because the impact may only manifest several years after its publication. In this case, they applied a weight to the publication journals according to the average number of citations received by the work. For this, the impact factor of the journal is used. The top 10 journals that have published the most papers (2019–2021) in the field of renewable energy in the agriculture sector are presented in Table 3.

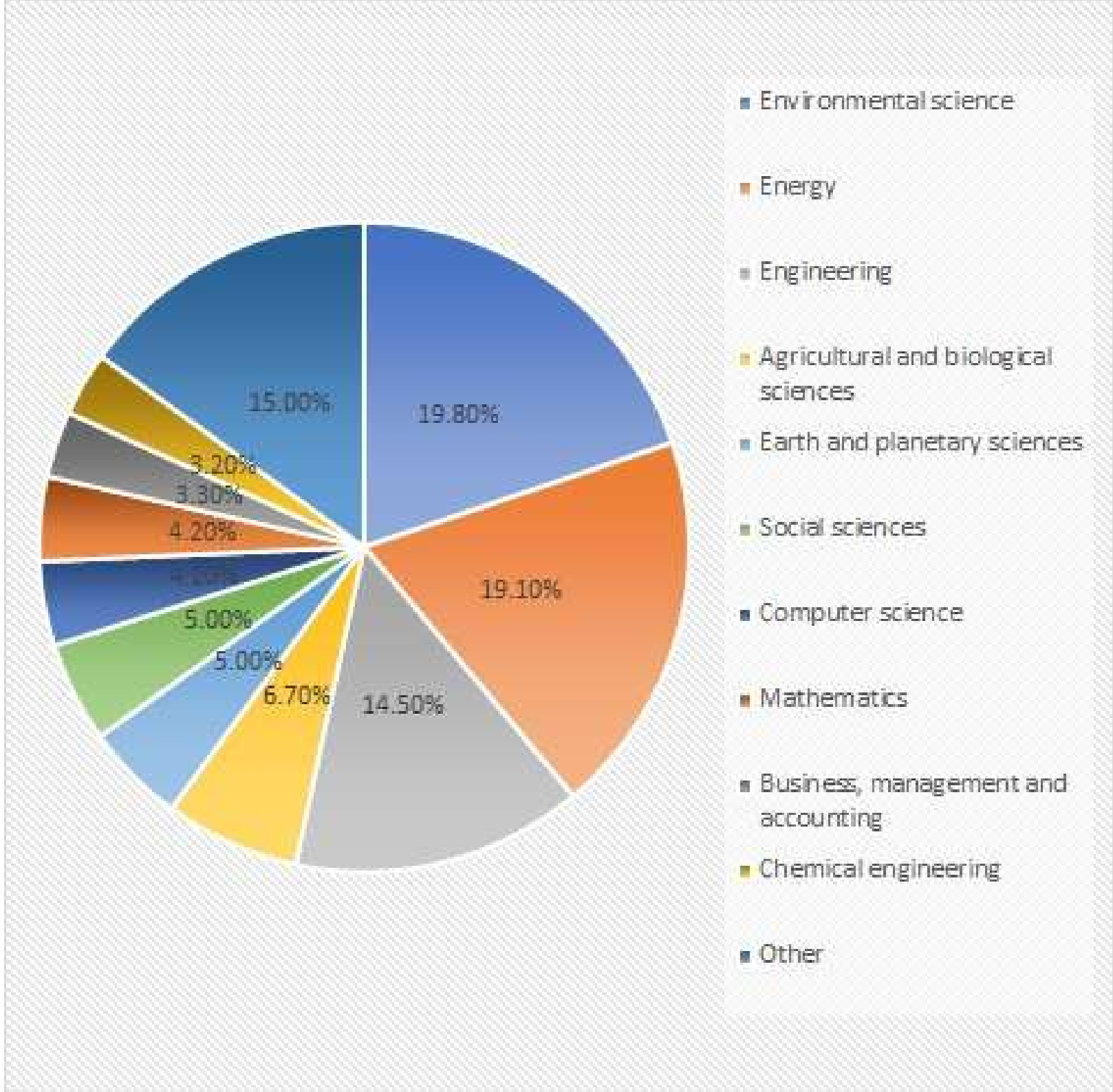

**Figure 7.** Documents published by subject area.

**Table 3.** Top 10 journals in 2021 in terms of documents published on renewable energy and agriculture.

| Journal | Quartile | SRJ | H-Index | JCR | Total Documents | Total Cites | Country |
|---|---|---|---|---|---|---|---|
| *Energies* | Q1 (SJR) | 0.653 | 111 | 3.252 | 57 | 76,498 | Switzerland |
| *Journal of Cleaner Production* | Q1 (SJR) | 1.921 | 232 | 11.072 | 51 | 250,231 | United Kingdom |
| *Renewable and Sustainable Energy Reviews* | Q1 (SJR) | 3.768 | 337 | 16.799 | 32 | 162,005 | United Kingdom |
| *Science of the Total Environment* | Q1 (SJR) | 1.806 | 275 | 10.754 | 31 | 328,230 | Netherlands |
| *Renewable Energy* | Q1 (SJR) | 1.877 | 210 | 8.634 | 30 | 95,713 | United Kingdom |
| *Sustainability* | Q1 (SJR) | 0.664 | 109 | 3.889 | 30 | 130,266 | Switzerland |
| *Environmental Science and Pollution Research* | Q1 (SJR) | 0.831 | 132 | 5.190 | 22 | 110,000 | Germany |
| *Applied Sciences* | Q2 (SJR) | 0.507 | 75 | 2.838 | 4 | 63,761 | Switzerland |
| *Clean Technologies and Environmental Policy* | Q1 (SJR) | 0.756 | 62 | 4.700 | 4 | 6258 | Germany |
| *Energy Reports* | Q1 (SJR) | 0.894 | 49 | 4.937 | 4 | 8369 | United Kingdom |

The three journals with the most documents published on the topic of renewable energy in the agriculture sector in the period of 2019–2021 are *Energies*, a journal with an H-Index of 111 and a JCR impact factor of 3.252, the Journal of Cleaner Production, with an H-Index of 232 and a JCR impact factor of 11.072, and Renewable and Sustainable Energy Reviews, a journal with an H-Index of 337 and a JCR impact factor of 16.799.

### *3.5. Most Cited Papers*

Table 4 shows the top 20 most cited documents.

**Table 4.** Top 20 most cited documents.

| Document Title | Year | Journal | Cited by | Reference |
|---|---|---|---|---|
| Effect of natural resources, renewable energy and economic development on $CO_2$ emissions in BRICS countries | 2019 | *Science of the Total Environment* | 338 | [39] |
| The 2021 report of the Lancet Countdown on health and climate change: code red for a healthy future | 2021 | *The Lancet* | 205 | [40] |
| Application of nanoparticles in biofuels: An overview | 2019 | *Fuel* | 162 | [41] |
| The greenhouse effect of the agriculture-economic growth-renewable energy nexus: Evidence from G20 countries | 2019 | *Science of the Total Environment* | 159 | [42] |
| Agrivoltaics provide mutual benefits across the food–energy–water nexus in drylands | 2019 | *Nature Sustainability* | 151 | [43] |
| The significance of biomass in a circular economy | 2020 | *Bioresource Technology* | 143 | [44] |
| Feasibility analysis and techno-economic design of grid-isolated hybrid renewable energy system for electrification of agriculture and irrigation area: A case study in Dongola, Sudan | 2019 | *Energy Conversion and Management* | 133 | [45] |
| Linking renewable energy, globalization, agriculture, $CO_2$ emissions and ecological footprint in BRIC countries: A sustainability perspective | 2021 | *Renewable Energy* | 125 | [46] |
| Agrophotovoltaic systems: applications, challenges, and opportunities. A review | 2019 | *Agronomy for Sustainable Development* | 119 | [47] |
| Energy and Food Security: Linkages through Price Volatility | 2019 | *Energy Policy* | 112 | [48] |
| Towards a sustainable environment: The nexus between ISO 14001, renewable energy consumption, access to electricity, agriculture and $CO_2$ emissions in SAARC countries | 2020 | *Sustainable Production and Consumption* | 107 | [49] |
| What abates ecological footprint in BRICS-T region? Exploring the influence of renewable energy, non-renewable energy, agriculture, forest area and financial development | 2021 | *Renewable Energy* | 105 | [50] |
| Effects of agriculture, renewable energy, and economic growth on carbon dioxide emissions: Evidence of the environmental Kuznets curve | 2020 | *Resources, Conservation and Recycling* | 100 | [51] |
| Evaluating the role of renewable energy, economic growth and agriculture on $CO_2$ emission in E7 countries | 2020 | *International Journal of Sustainable Energy* | 100 | [52] |
| Selecting sustainable energy conversion technologies for agricultural residues: A fuzzy AHP-VIKOR based prioritization from life cycle perspective | 2019 | *Resources, Conservation and Recycling* | 99 | [53] |
| Anaerobic digestion of livestock manure in cold regions: Technological advancements and global impacts | 2020 | *Renewable and Sustainable Energy Reviews* | 88 | [54] |
| Contribution to Circular Economy options of mixed agricultural wastes management: Coupling anaerobic digestion with gasification for enhanced energy and material recovery | 2019 | *Journal of Cleaner Production* | 88 | [55] |
| Small-scale urban agriculture results in high yields but requires judicious management of inputs to achieve sustainability | 2019 | *Proceedings of the National Academy of Sciences of the United States of America* | 88 | [56] |
| Revisiting the role of forestry, agriculture, and renewable energy in testing environment Kuznets curve in Pakistan: evidence from Quantile ARDL approach | 2020 | *Environmental Science and Pollution Research* | 86 | [57] |
| Exploring the nexus between agriculture and greenhouse gas emissions in BIMSTEC region: The role of renewable energy and human capital as moderators | 2021 | *Journal of Environmental Management* | 82 | [58] |

In their report issued in 2021, Romanello et al. [40] state that the world has made little progress in controlling climate change, causing, among other things, environmental suitability for the transmission of infectious diseases, extreme weather events, and worsened emotional and physical well-being.

The environmental Kuznets curve states that there is an inverted U-shaped relationship between economic growth and environmental degradation in such a way that environmental degradation increases as the level of wealth of an economy increases up to a maximum threshold, after which, even as the wealth of an economy increases, environmental degradation decreases. Within this framework, several studies have been carried out in countries throughout the world. Danish et al. [39] considered natural resources, economic development, renewable energies, and $CO_2$ emissions in Brazil, China, Russia, and South Africa as factors for their study. Quiao et al. [42] estimated agriculture, economic growth, and renewable energies to be factors causing carbon dioxide emissions in G20 countries. Pata [46] performed statistical tests to analyze the effect of renewable energy generation, globalization, and agricultural activities on the ecological footprint and $CO_2$ emissions of BRIC countries between the years 1971 and 2016. In the same vein, Aziz et al. [57] conducted a study in Pakistan, Sharma et al. [58] conducted a study of the countries grouped in the international organization BIMSTEC, Ridzuan et al. [51] conducted a study in Malaysia, and Aydoğan et al. [52] conducted a study using a panel of E7 countries.

Sherwood [44] reviewed biomass production and feedstock use by implementing the circular economy framework from a European perspective, concluding that it is important to decouple the petrochemical industry and biomass production in order to minimize waste from food and agriculture and to encourage better cooperation between all actors in the value chain. Taghizadeh-Hesary et al. [48] highlighted the significant impact of oil prices on food prices and concluded that diversifying energy consumption in the agri-food sector by using renewable resources is necessary. Ikram et al. [49] investigated the relationship between ISO 14001, renewable energy consumption, access to electricity, agriculture, and $CO_2$ emissions in South Asian partnership countries. Usman and Makhdum [50] studied the influence of agriculture, forestry, renewable energy use, and financial development on ecological footprints in Brazil, Russia, India, China, and Turkey.

The use of technologies applied to renewable energy production is a determining factor in the formation of the environmental Kuznets curve. Sekoai et al. [41] reviewed publications related to the use of nanoparticles in improving the performance of biofuel production processes. Barron-Gafford et al. [43] and Weseleck et al. [47] studied agrovoltaic systems that led to improved crop yields and better soil utilization of water resources, while Yao et al. [54] reviewed articles that involved anaerobic digestion of manure for biogas utilization. Antoniou et al. [55] used digestion for biogas transformation and Wang et al. [53] concluded that, in China, direct combustion energy production, gasification energy production, and briquette fuel are the most environmentally and economically sustainable technologies. Elkadeem et al. [45] studied the possibility of installing a clean and sustainable off-grid energy system in a region of Dongola in Sudan. Among the different combinations of photovoltaic, wind turbine, diesel generator, battery storage, and converter systems, the solar-powered hybrid system managed to best meet the established demand, reducing the cost of energy production and obtaining a positive return on investment. McDougall et al. [56] studied urban crops as an alternative to traditional crops. The authors concluded that these crops improved land use; however, care had to be taken to ensure sustainability.

### *3.6. Scientific Community and Keyword Analysis*

Identifying the most prominent authors on renewable energy applied in the agriculture sector and their most used keywords will help in understanding the scientific research trends for this subject.

Table 5 presents information about the most productive authors, such as the number of documents in the last three years, affiliation, H-Index, and country.

**Table 5.** Information about the top 5 authors.

| Author | No. of Documents in the Last Three Years | Affiliation | H-Index | Country | Investigation Line |
|---|---|---|---|---|---|
| Zhao, L. | 6 | Ministry of Agriculture of the People's Republic of China | 23 | China | Agricultural biomass for energy |
| Huo, L. | 5 | Ministry of Agriculture of the People's Republic of China | 12 | China | Agricultural biomass for energy |
| Kodirov, D. | 5 | Tashkent Institute of Irrigation and Agricultural Mechanization Engineers at the National Research University of Uzbekistan | 9 | Uzbekistan | Renewable energy technologies in the agricultural sector |
| Yao, Z. | 5 | Chinese Academy of Agricultural Sciences | 17 | China | Agricultural biomass for energy |
| Daoutidis, P. | 4 | University of Minnesota, Twin Cities | 11 | USA | Renewable energy systems in agriculture |

Identifying groups of scientific communities helps identify current research, the quantity of relations between authors from different institutions, and areas of knowledge [59]. The identification of communities has been applied in energy [60] and education [28] with great success. In Figure 8, a map of scientific communities made using VOSviewer software version 1.6.18 is presented. The color assigned to each author determines which cluster it belongs to. The circle's size indicates the author's reputation based on the number of publications on renewable energy applied in agriculture. The lines represent the collaborations between the different authors that form each cluster.

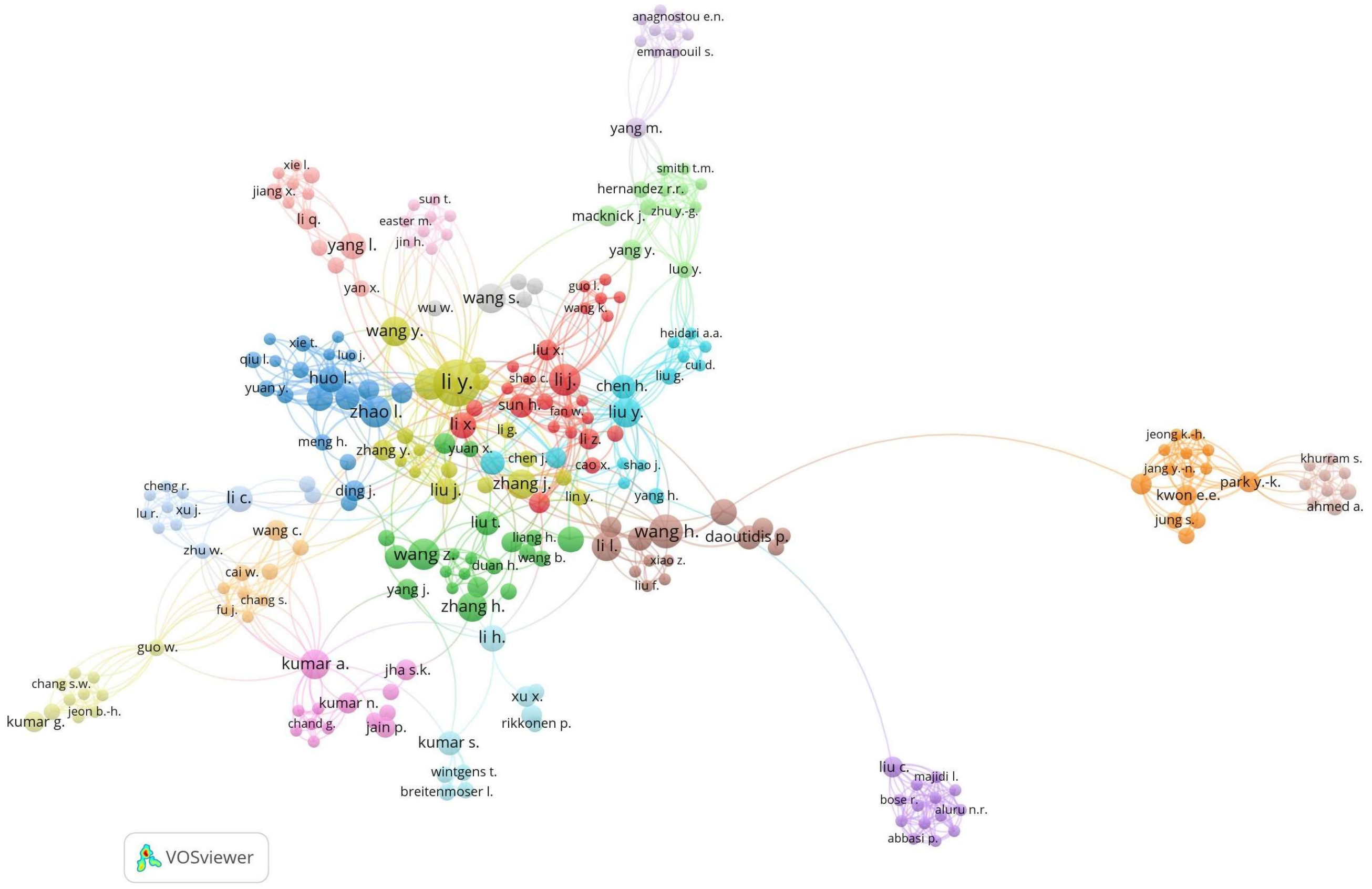

**Figure 8.** International scientific clustering.

An analysis of the most used keywords in all publications on renewable energy applied in agriculture that were published between 2019 and 2021 was carried out in this work. Figure 9 illustrates the keywords most used by the authors in their research on renewable energies in the agricultural sector, where the size of the keyword is related to the frequency of occurrence of that keyword in the scientific articles analyzed.

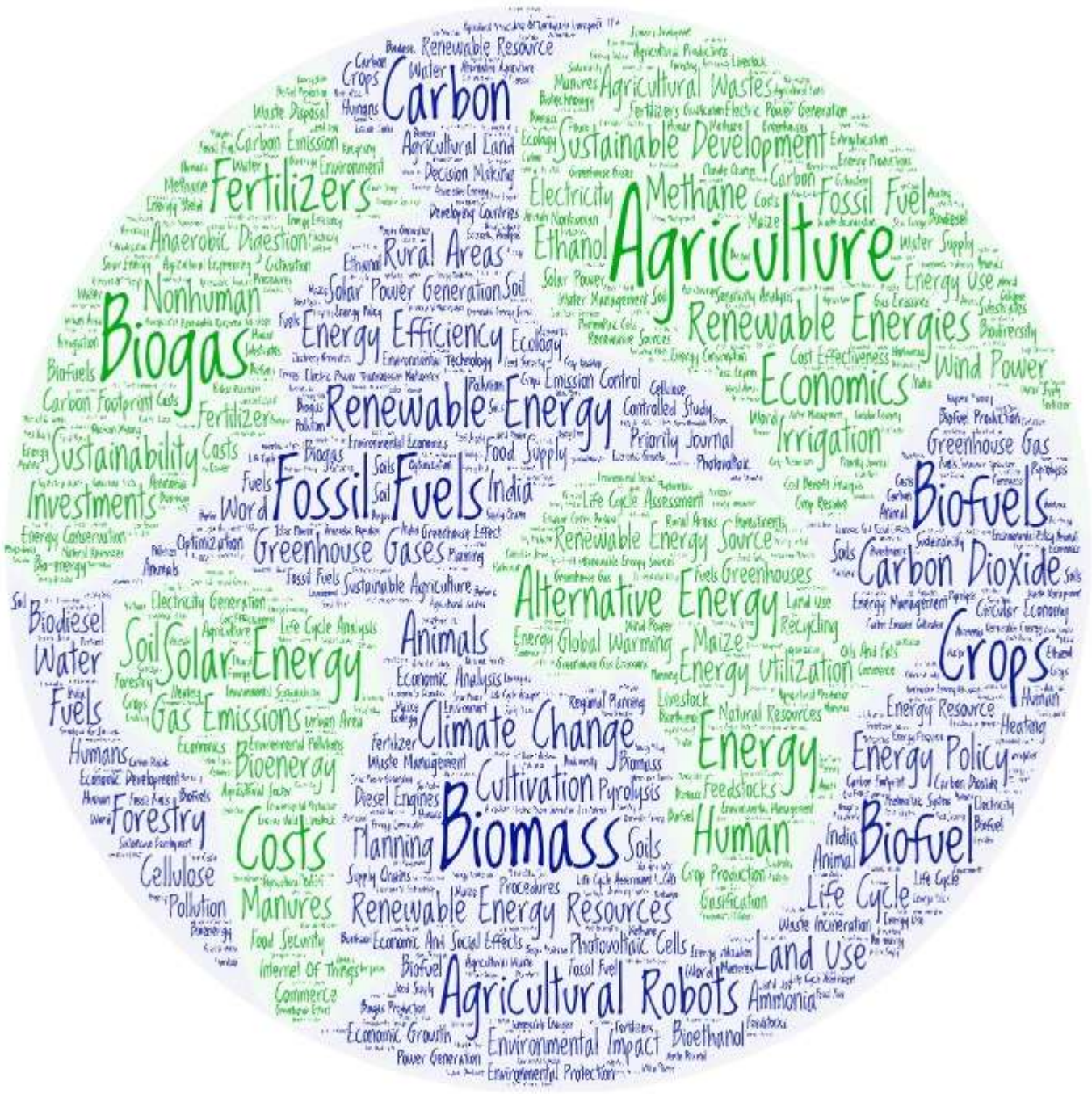

**Figure 9.** Keywords used in research on renewable energy applied in agriculture.

Keywords are the main instrument of research studies. They are very useful for bibliometric analysis and for fine-tuning search equations, as well as for analyzing trends in a given line of research.

Specifically, in this study, the keyword "Agriculture" was found to have been mentioned 658 times and was the most mentioned, followed by "Agricultural robots", "Renewable Energy", "Biomass", and "sustainable development"; with this analysis it is possible to determine research trends.

As a result of the visualization of keywords in the bibliometric map (Figure 10), five large thematic groups (clusters) were obtained, which suggest the main emerging research focuses in the area. Table 6 shows the color, main words, and topic for each cluster.

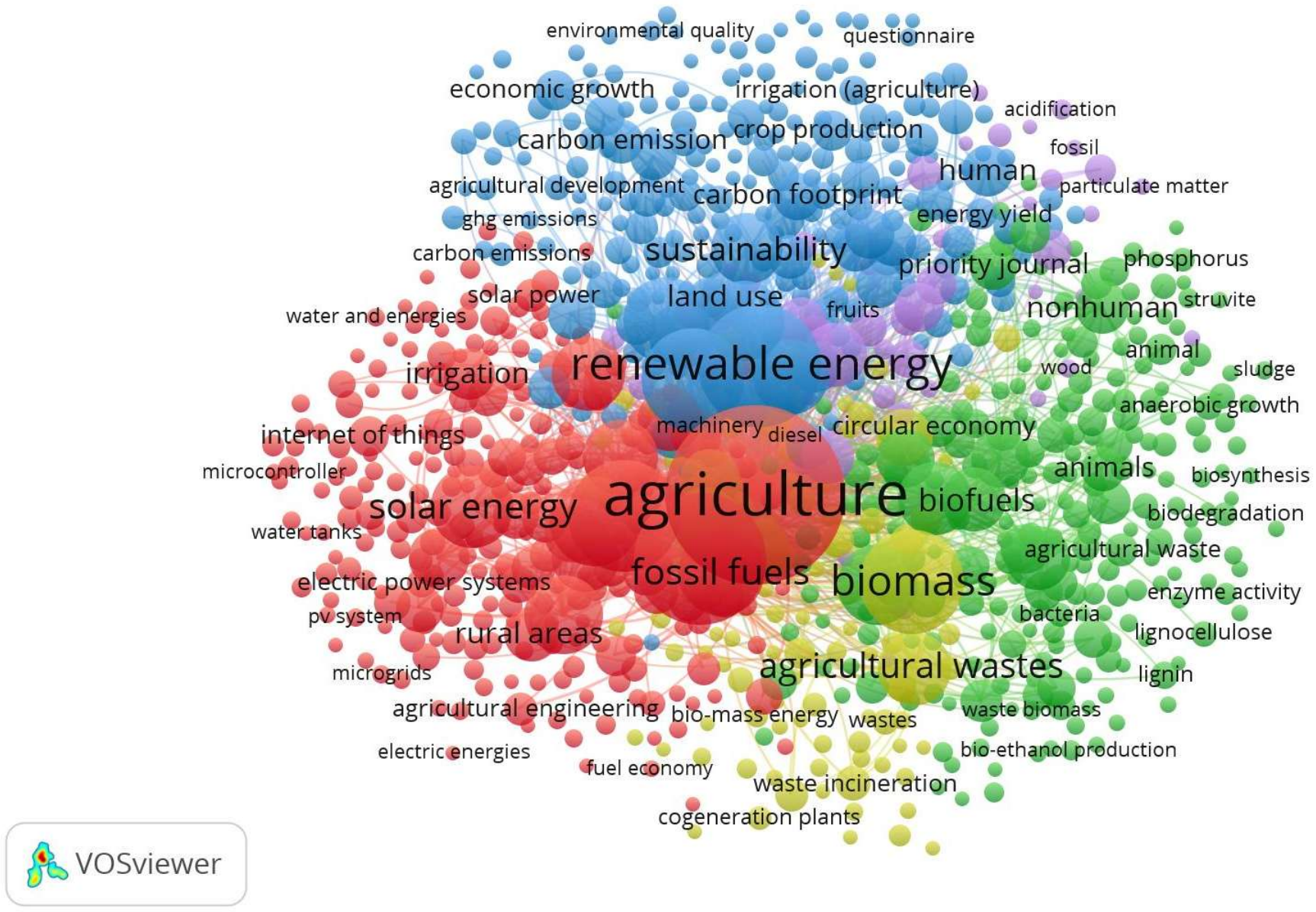

**Figure 10.** Analysis of the co-occurrence of words.

**Table 6.** Main clusters obtained in co-occurrence analysis for the topic of renewable energies in agriculture.

| Cluster | Color | Main Keywords | Topic |
|---|---|---|---|
| 1 | Red | Agriculture Robots, Solar Energy, Renewable Energy Sources, Energy Utilization, Irrigation. | Agriculture–Renewable energy |
| 2 | Green | Biogas, Waste Management, Pyrolysis, Biofuel, Bioconversion. | Bioenergy |
| 3 | Blue | Renewable Energy, Alternative Energy, Sustainability, Crop Production, Carbon Dioxide. | Sustainable Agriculture |
| 4 | Yellow | Biomass, Agricultural Wastes, Bioenergy, Gas Emissions, Biomass Power. | Biomass Energy |
| 5 | Purple | Environmental Impact, Global Warming, Life Cycle, Greenhouse Gas Emissions, Fertilizer. | Environmental Impact of Agriculture |

## 4. Conclusions

In this study, scientific publications related to renewable energies in the agricultural sector that were published in the last three years were analyzed.

Analysis of data obtained from the Scopus database leads us to conclude that studies regarding renewable energy applied in the agricultural sector greatly increased in the three years analyzed. It can be seen that many of the studies carried out in recent years have been a great step toward the integration of renewable energies in the agricultural sector and are the result of the end of the first cycle of commitments made in the Kyoto Protocol.

The three main categories in which these research papers were grouped were environmental science (598 documents, equivalent to 19.8% of the total documents published), energy (19.1%), and engineering (14.5%).

The top five institutions carrying out research in this area include three from China (Chinese Ministry of Education, Chinese Agricultural University, and Chinese Academy of Sciences), one from Iran (University of Tehran), and one from the Netherlands (Wageningen University and Research). The main countries that contributed to the research were India, China, the United States, Italy, and the United Kingdom; these countries were found in the leading clusters.

The three principal journals with the highest number of publications were the *Energies* journal (with 57 documents published, an impact factor of 3.252, and an H-Index of 111), the *Journal of Cleaner Production* (with an H-Index of 232, 51 documents published, and a JCR impact factor of 11.072), and the *Renewable and Sustainable Energy Reviews* journal (33 documents published, the highest impact factor (16.799), and the highest H-Index (337)).

Keywords were used to identify scientific communities worldwide and their collaboration with other groups. In this study, the keyword "Agriculture" was found to have been mentioned most with 658 occurrences, followed by "Agricultural robots", "Renewable Energy", "Biomass", and "sustainable development"; with this analysis it is possible to determine research trends. Five clusters were found, which focused on renewable energy technologies in agriculture, bioenergy, sustainable agriculture, biomass energy, and the environmental impact of agriculture.

Collaboration between countries showed that the more important clusters were from the US, China, the United Kingdom, and Italy, countries with more influence and economic resources.

The most productive authors for this topic had the following investigation lines: agricultural biomass for energy, renewable energy technologies in the agricultural sector, and renewable energy systems in agriculture.

This study is expected to be extended in the future to other sectors of great importance to renewable energies and those which contribute greatly to their development.

**Author Contributions:** Conceptualization, Q.H.-E., D.M.-R., A.V.-C., J.M.J.L., P.A.-M., M.P.M.-J., and A.-J.P.-M.; methodology, Q.H.-E., D.M.-R., A.V.-C., J.M.J.L., P.A.-M., M.P.M.-J., and A.-J.P.-M.; investigation, Q.H.-E., D.M.-R., A.V.-C., J.M.J.L., P.A.-M., M.P.M.-J., and A.-J.P.-M.; writing—original draft preparation, Q.H.-E., D.M.-R., A.V.-C., J.M.J.L., P.A.-M., M.P.M.-J., and A.-J.P.-M. All authors have read and agreed to the published version of the manuscript.

**Funding:** This research received no external funding.

**Data Availability Statement:** Not applicable.

**Conflicts of Interest:** The authors declare no conflict of interest.

1.